\documentclass[useAMS,usenatbib]{mn2e}
\usepackage{epsfig}
\usepackage{color}

\title[Variation of microphysics in wind bubbles]{Variation of microphysics in wind bubbles: an alternative
mechanism for explaining the rebrightenings in GRB afterglows}
\author[S. W. Kong, A. Y. L. Wong, Y. F. Huang and K. S. Cheng]{S. W. Kong$^{1,2}$, A. Y. L.
Wong$^{2}$, Y. F. Huang$^{1}$ and K. S. Cheng$^{2}$\thanks{E-mail: hrspksc@hkucc.hku.hk (KSC)}\\
$^{1}$Department of Astronomy, Nanjing University,
Nanjing 210093, China\\
$^{2}$Department of Physics, The University of Hong Kong, Pokfulam
Road, Hong Kong, China}
\begin{document}

\date{Accepted ? December ?. Received ? December ?; in original form ? October ?}

\pagerange{\pageref{firstpage}--\pageref{lastpage}} \pubyear{2009}

\maketitle

\label{firstpage}

\begin{abstract}
Conventionally, long GRBs are thought to be caused by the core
collapses of massive stars. During the lifetime of a massive star, a
stellar wind bubble environment should be produced. Furthermore, the
microphysics shock parameters may vary along with the evolution of
the fireball. Here we investigate the variation of the microphysics
shock parameters under the condition of wind bubble environment,
\textbf{and allow the microphysics shock parameters to be
discontinuous at shocks in the ambient medium. It is found that our
model can acceptably reproduce the rebrightenings observed in GRB
afterglows, at least in some cases.} The effects of various model
parameters on rebrightenings are investigated. The rebrightenings
observed in both the R-band and X-ray afterglow light curves of GRB
060206, GRB 070311 and GRB 071010A are reproduced in this model.
\end{abstract}


\section{Introduction}

Gamma-ray bursts (GRBs) are attractive astrophysics phenomena and
had puzzled astronomers for about twenty-four years after their
discovery in 1973 (Klebesadel et al. 1973). The discovery of
long-lived, multi-band counterparts of GRBs, namely, the afterglows
of GRBs, in 1997 is a watershed in GRB research (Costa et al. 1997;
Van Paradijs et al. 1997; Frail et al. 1997). Soon after that, the
so-called fireball model is recognized as the standard model in
view of the fact that it can explain most features of GRB
observations well. However, as the advance of observation techniques
and the accumulation of observational data, especially after the
launch of \emph{Swift} satellite (Gehrels 2004), a lot of unexpected
behaviours appear in GRB afterglows, such as the canonical
steep-shallow-normal decay and flares in X-ray afterglows, the
flattish decay phase and various rebrightenings in optical
afterglows (For review, see Zhang 2007).

In fact, there are more and more rebrightenings detected in GRB
afterglows, including GRB 970508 (Galama et al. 1998a), GRB 990123
(Sari \& Piran 1999), GRB 021004 (Lazzati et al. 2002), GRB 030329
(Berger et al. 2003), GRB 050525A (Blustin et al. 2006), GRB 050820A
(Cenko et al. 2006), GRB 050721 (Antonelli et al. 2006), GRB 060206
(Stanek et al. 2007; Liu et al. 2008), GRB 070125 (Updike et al.
2008), GRB 070311 (Guidorzi et al. 2007a), GRB 071003 (Perley et al.
2008), GRB 071010A (Covino et al. 2008). Many different mechanisms
have been proposed to explain these rebrightenings, such as density
jump (Lazzati et al. 2002; Dai \& Wu 2003; Tam et al. 2005), energy
injection (Huang et al. 2006), two-component jet (Huang et al. 2004,
2006; Liu et al. 2008), reverse shock (Sari \& Piran 1999),
reverberation of the energy input measured by prompt emission
(Vestrand et al. 2006), turn-on of the external shock (Stanek et al.
2007; Molinari et al. 2007), large angle emission (Panaitescu \&
Kumar 2007), and spectral peak of existing forward shock (Shao \&
Dai 2005).

Among all the mechanisms, the density jump model needs to be paid
special attention to, because density jump in surrounding medium of
GRBs is a very natural hypothesis. Since there are more and more
examples indicating that some GRBs are associated with Type Ic
supernovae (e.g. SN 1998bw/GRB980425, Galama et al. 1998b;
SN2003dh/GRB030329, Price et al. 2003) and many host galaxies of
GRBs are in process of active star formation (Fruchter et al. 1999;
Djorgovski et al. 1998), it is believed that the progenitors of long
GRBs are massive Wolf-Rayet (WR) stars (Woosley 1993). Massive stars
usually produce very strong stellar wind to push the initial
interstellar medium (ISM) in their neighborhood away during their
lifetime. The surrounding of these GRBs should be a wind bubble
following the density profile $\rho \propto r^{-2}$ rather than the
usual homogeneous ISM. Several authors (Castor et al. 1975; Weaver
1977; Ramirez-Riuz et al. 2001) further found that beyond some
typical radius of the wind bubble, the swept-up mass is too large to
be pushed by the wind. As a result, the wind materials pile up at
the edge of the wind bubble to form a density jump.

Lazzati et al. (2002) proposed that a density jump can lead to a
rebrightening in the afterglow. In usual case, when the observing
frequence is between the peak frequence ($\nu_{\rm m}$) and the
cooling frequence ($\nu_{\rm c}$), the amplitude of the
rebrightening should be proportional to the square root of the
density contrast. However, if the density contrast is too high, the
rebrightening will be weakened, since $\nu_{\rm c}$ will decrease
and become less than the observing frequence. In the more detailed
numerical simulations by Huang et al. (2007), no obvious
rebrightenings can be discriminated when the density jump amplitude
is set to 100, which suggests that the simple density jump model is
not an ideal mechanism to produce the rebrightenings. In the recent
studies by Nakar \& Granot (2007) and van Eerten et al. (2009), a
full treatment of the transient features at the jump moment, even
including the thickness of the blastwave, was considered. Thus these
studies should be a more accurate approximation to the reality.
Interestingly, no obvious rebrightenings associated with the density
jumps are found.

In short, a simple density jump model is difficult to explain the
observed rebrightenings in GRB afterglows. On the other hand, the
early afterglows of most GRBs exhibit flattish decays with $\alpha <
0.8$, where $\alpha$ is defined as $F_{\nu} \propto t^{-\alpha}$.
This is difficult to explain using the standard fireball model when
the density profile is $\rho \propto r^{-2}$. Some previous works
have suggested that the microphysics parameters may vary during the
evolution of the fireball (Rossi \& Rees 2003; Ioka et al. 2006; Fan
\& Piran 2006; Panaitescu et al. 2006; Granot, K\"{o}nigl \& Piran
2006). We believe that this kind of variation could resolve these
problems.

In this paper, we show that the observed rebrightenings in GRB
afterglows can be well reproduced by assuming varying microphysics
shock parameters associated with the wind bubble environments. The
outline of our paper is as follows: in \S 2 we introduce our model
detailedly. We then numerically investigate the effects of various
parameters on the afterglows in \S 3, and reproduce the R-band and
X-ray afterglow light curves of GRB 060206, GRB 070311 and GRB
071010A in \S 4. Our discussion and conclusions are presented in \S
5. We use an assumptive cosmology of $H_{\rm 0} = 65$ ${\rm km}$
${\rm s}^{-1}$ ${\rm Mpc}^{-1}$, $\Omega_{\rm M} = 0.30$ and
$\Omega_{\rm \Lambda} = 0.70$ throughout the paper.

\section{Model}

\subsection{Dynamics and radiation process}

In the standard fireball model, the outflow of GRB, which moves
relativistically, interacts with the surrounding medium to form an
external shock. A constant fraction $\xi_{\rm e}$ of the shock
energy will be transferred to the swept-up electrons and accelerate
them to relativistic velocity. Similarly, a constant fraction
$\xi_{\rm B}$ of the shock energy will go to the magnetic field.
These shocked relativistic electrons move in the magnetic field and
emit synchrotron radiation to produce broadband afterglows.

We use the convenient equations developed by Huang et al. (1999,
2000a, 2000b, 2003) to describe the dynamics and radiation process
of the ejecta. The evolution of the bulk Lorentz factor $\gamma$,
the shock radius $R$, and the swept-up medium mass $m$, is described
by three differential equations,
\begin{equation}
\frac{{\rm d} \gamma}{{\rm d} m} = - \frac{\gamma^2 - 1}
       {M_{\rm ej} + \epsilon m + 2 ( 1 - \epsilon) \gamma m},
\end{equation}
\begin{equation}
\frac{{\rm d} R}{{\rm d} t} = \beta c \gamma (\gamma +
\sqrt{\gamma^2 - 1}),
\end{equation}
\begin{equation}
\frac{{\rm d} m}{{\rm d} R} = 2 \pi R^2 (1 - {\rm \cos} \theta) n
m_p,
\end{equation}
where $m_p$ is proton mass, $M_{\rm ej}$ is the initial mass of the
outflow, $\theta$ is the half opening angle of the jet, $n$ is the
number density of the environment, $\beta = \sqrt{\gamma^2 -
1}/\gamma$,  and $\hat{\gamma} \approx (4 \gamma + 1)/(3 \gamma)$ is
the adiabatic index (Dai, Huang, \& Lu 1999). $\epsilon$ is the
radiative efficiency, which equals 1 for a highly radiative case,
and equals 0 in the adiabatic case. We ignore the sideways expansion
of the jet in our model, because many numerical simulations indicate
that it is a very slow process (Granot et al. 2001; Cannizzo,
Gehrels \& Vishniac 2004; Zhang \& MacFadyen 2009). We consider that
the afterglow flux mainly comes from the synchrotron radiation of
the shocked relativistic electrons.

Since the speed of light ( $c$ ) is finite, photons received by the
observer at a particular time $t$ are not radiated simultaneously,
but come from a distorted ellipsoid determined by
\begin{equation}
t=\int \frac{1 - \beta {\rm \cos} \Theta}{\beta c} {\rm d} R \equiv
{\rm const},
\end{equation}
within the jet boundaries, where $\Theta$ is the angle between the
velocity of emitting material and the line of sight. This is the
so-called EATS effect. We integrate
over the EATS to obtain an accurate observed flux in our
simulations (see Huang et al. 2007 for more details).

\subsection{Environment}

Massive stars usually produce very strong stellar wind during their
lifetime. This stellar wind interacts with the initial ISM and forms
two shocks: a reverse shock propagates back into the stellar wind
and a forward shock that propagates into the ISM. The resulting
surrounding medium is broken into four regions (Castor et al. 1975;
Weaver 1977; Ramirez-Riuz et al. 2005; Pe'er and Wijers 2006) as
shown in Figure~1. They are (from inside to outside):
(1) the unshocked stellar wind;
(2) the shocked stellar wind; (3) the shocked ISM; (4) the unshocked
ISM.

\begin{figure}
\resizebox{\hsize}{!}{\includegraphics{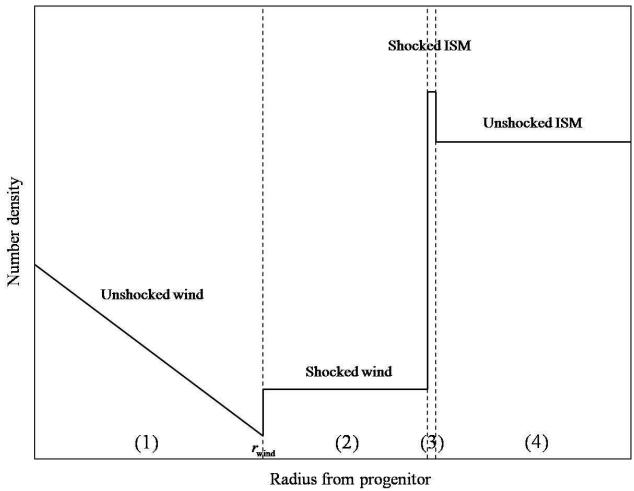}} \caption{The
illustration of the density profile used in our model.}
\end{figure}

Region (1) is a stellar wind bubble. The number density of this
region is given by
\begin{equation}
n_1=\frac{\dot M}{4 \pi m_p v_{\rm w} r^2} \propto r^{-2},
\end{equation}
where $\dot M$ is the mass loss rate of the progenitor, $v_{\rm w}$
is the launching speed of the stellar wind and $r$ is the radial
distance from the central star. We fix $v_{\rm w}$ as $10^8$ cm ${\rm
s}^{-1}$ and only vary $\dot M$ in our simulations, since
the density profile of the environment only depends on the
ratio $\dot M/v_{\rm w}$.

As the stellar wind expands, the swept-up mass of the ISM increases
and is finally comparable to the mass in the wind at a radius
$r_{\rm wind}$, which is determined by (Ramirez-Ruiz et al. 2001,
2005)
\begin{equation}
r_{\rm wind}=1.6 \times 10^{18} {\dot M}_{-6}^{3/10} n_{{\rm
ISM},3}^{-3/10} v_{{\rm w},8}^{1/10} t_{\star,6}^{2/5} \rm cm,
\end{equation}
where $n_{\rm ISM,3}$ is the initial density of the homogeneous ISM,
in units of $10^3$ ${\rm cm}^{-3}$, and $t_{\star,6}$
is the life time of the WR phase of the star, in units of $10^6$ yr. Beyond
this radius, in region (2), the swept-up mass is too large to be
pushed away by the wind. The material piles up in this region to form a
density jump. Because of the effect of the reverse wind shock, the
material is hot in this region. The density is approximately
constant and equals to $n_2 = 4n_1(r_{\rm wind})$, where $n_1(r_{\rm
wind})$ is the density in region (1) at the radius of $r = r_{\rm
wind}$.

As shown by Castor et al. (1975), the boundary between Region (2) and
Region (3) is at the radius of
\begin{equation}
r_{2-3}=1.6 \times 10^{19} {\dot M}_{-6}^{1/5} n_{{\rm
ISM},3}^{-1/5} v_{{\rm w},8}^{2/5} t_{\star,6}^{3/5} \rm cm.
\end{equation}
Region (3) is a thin, dense shell containing most of the swept-up
ISM. Its width is only about $r_{2-3}/12$ and the density in this region
is about $4n_{\rm ISM}$. Here $n_{\rm ISM}$ is also the density in region
(4).

We use the density profile introduced above as the environment
surrounding GRBs in this paper. As discussed by Pe'er and Wijers
(2006), for a GRB event, the blast wave can not reach Region (3) in
the relativistic phase. Our numerical results also prove that the
fireball is in Region (1) and Region (2) during typical observable
time. So we only use Region (1) and Region (2) as the environment in
our work. For simplicity, we take $r_{\rm wind}$ as a free parameter
in our calculations, so that we do not need to consider the detailed
values of $v_{\rm w}$ and $t_{\star}$.

Unfortunately, the simple density jump model alone is difficult to explain
the rebrightenings in GRB afterglows (Huang et al. 2006, 2007; Nakar \&
Granot 2007). Moreover, the early optical and/or X-ray
afterglow light curves of some GRBs are flatter than the prediction of a
normal fireball in a simple wind environment. So we need to
consider other effects. Varying microphysics shock parameters should be
a possible solution.

\subsection{Microphysics parameters}

In the standard afterglow model, it is usually assumed that the
microphysical parameters are constant throughout the evolution of
the fireball. However, the actual microphysical processes in the
relativistic shocks, for example, the energy transfer from protons
to electrons and magnetic fields, are still not clearly known. It is
possible that the microphysical parameters are variational. In fact,
the variation of the microphysics parameters during the evolution of
the fireball has been considered in many previous studies (Rossi \&
Rees 2003; Ioka et al. 2006; Fan \& Piran 2006; Panaitescu et al.
2006; Granot, K\"{o}nigl \& Piran 2006). Fan \& Piran (2006) and
Panaitescu et al. (2006) have engaged the assumption that the
equipartition factors $\xi_{\rm e}$ and $\xi_{\rm B}$ are functions
of $\gamma$ to explain the shallow decay phase in some X-ray
afterglows. Here we will use a similar idea to investigate the
observed rebrightenings in GRB afterglows.

In our model, the circum-burst environment is divided into two parts,
i.e. Region (1) and Region (2). The material in Region (1) is the
unshocked stellar wind thrown by the mass star. In
Region (2) the material consists of the shocked stellar wind mixed
with a small fraction of the swept-up ISM. It is hot because of the
effect of the reverse wind shock (Castor et al. 1975; Weaver 1977;
Ramirez-Riuz et al. 2005; Pe'er and Wijers 2006). We can imagine
that the physical condition, such as the strength of the magnetic
field, the temperature and density of the material, could be different
between these two regions, so the evolution of microphysics
parameters may also be different accordingly. In our study, we use different
parameters for these two regions. We assume that
\begin{equation}
\xi_e=\xi_{e,0} \gamma^{-\alpha1},
\end{equation}
\begin{equation}
\xi_B=\xi_{B,0} \gamma^{-\alpha1}
\end{equation}
in Region (1), and
\begin{equation}
\xi_e=\xi_{e,0} \gamma^{-\alpha2},
\end{equation}
\begin{equation}
\xi_B=\xi_{B,0} \gamma^{-\alpha2}
\end{equation}
in Region (2).

\section{Numerical Results}

In this section, we present our numerical results to show the
effects of various physical parameters on the R-band light curves,
using the model described in \S 2. For convenience, we first define
a set of ``standard'' parameters, as shown in the second column
of Table 1.
\begin{table*}
 \centering
 \begin{minipage}{100mm}
  \caption{Modeling Parameters}
  \begin{tabular}{c c c c c}
\hline \hline Parameters & Standard & GRB 060206 & GRB 070311 & GRB 071010A\\
\hline
$z$ & 1.0 & 4.045 & 1.0 & 0.98\\
$E_0$ (ergs) & $1.0 \times 10^{53}$ & $9.0\times 10^{52}$ & $8.0 \times 10^{51}$ & $1.3 \times 10^{52}$\\
$\theta_0$ (rad) & 0.1 & 0.06 & 0.08 & 0.11\\
$\dot M$ ($M_\odot/{\rm yr}$) & $1.0 \times 10^{-6}$ & $2.0 \times 10^{-5}$ & $6.0 \times 10^{-6}$ & $4.0 \times 10^{-5}$\\
$r_{\rm wind}$ (${\rm cm}$) & $1.0 \times 10^{18}$ & 3$.0 \times 10^{16}$ & $2.0 \times 10^{17}$ & $5.1 \times 10^{16}$\\
$p$ & 2.5 & 2.2 & 2.3 & 2.1\\
$\xi_{e,0}$ & 0.1 & 0.3 & 0.3 & 0.3\\
$\xi_{B,0}$ & 0.01 & 0.03 & 0.03 & 0.03\\
$\alpha1$ & 1.5 & 1.2 & 0.9 & 1.5\\
$\alpha2$ & 1.3 & 0.6 & 0.0 & 0.6\\
\hline
\end{tabular}
\end{minipage}
\end{table*}

Figure 2 illustrates the effect of the parameter $\theta_0$ on the
R-band light curves. The solid line corresponds to the ``standard''
parameters. The dashed line corresponds to $\theta_0 = 0.3$ rad and
the dotted line corresponds to $\theta_0 = 0.03$ rad. We can see
that when the change of microphysics parameters is considered, an
obvious rebrightening appears. Interestingly, when $\theta_0$ is
larger, the duration of the rebrightening becomes longer. This is
not difficult to understand. The afterglow brightness is dominated
by the high latitude emission, so that the EATS shows a ring-like
pattern. As a result, the brightness will be kept on a relatively
higher level after the density jump moment, to form a flat phase.
When $\theta_0$ becomes larger, the time delay of the photons from
the edge of the outflow also becomes larger. So the flat phase is
longer.

\begin{figure}
\resizebox{\hsize}{!}{\includegraphics{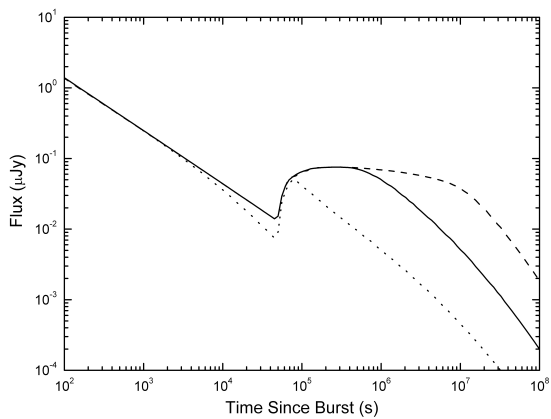}} \caption{The effect
of the parameter $\theta_0$ on the R-band light curve. The solid
line corresponds to the standard parameters. The dashed line
corresponds to $\theta_0 = 0.3$ rad and the dotted line corresponds
to $\theta_0 = 0.03$ rad.}
\end{figure}

Environment is an important factor that affects the evolution of GRB
afterglows. As mentioned before, the character of wind environment
is mainly determined by the ratio $\dot M/v_{\rm w}$. We fix $v_{\rm
w}$ as $10^8$ cm ${\rm s}^{-1}$ and only vary $\dot M$ here to
investigate the effect of the parameter $\dot M$ on the R-band light
curves. The results are shown in Figure 3. The solid line
corresponds to the ``standard'' parameters. The dashed line
corresponds to ${\dot M} = 1.0 \times 10^{-5}$ $M_\odot/{\rm yr}$
and the dotted line corresponds to ${\dot M} = 1.0 \times 10^{-7}$
$M_\odot/{\rm yr}$. It is clearly seen that for a smaller $\dot M$
value, the rebrightening appears earlier and the flux before the
rebrightening are lower. This result is easy to understand. A
smaller $\dot M$ corresponds to a smaller density, and the
deceleration of the external shock is slower. So the Lorentz factor
is larger and the fireball can meet the density jump earlier. A
larger Lorentz factor also makes $\xi_{\rm e}$ and $\xi_{\rm B}$
smaller and decreases the flux.
\begin{figure}
\resizebox{\hsize}{!}{\includegraphics{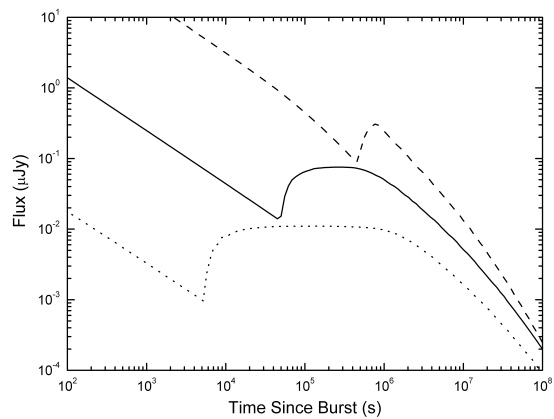}} \caption{The effect
of the parameter $\dot M$ on the R-band light curve. The solid line
corresponds to the ``standard'' parameters. The dashed line corresponds
to ${\dot M} = 1.0 \times 10^{-5}$ $M_\odot/{\rm yr}$ and the dotted
line corresponds to ${\dot M} = 1.0 \times 10^{-7}$ $M_\odot/{\rm
yr}$.}
\end{figure}

The parameter $r_{\rm wind}$ is another factor to describe the
character of the environment. The effect of $r_{\rm wind}$ on the
R-band light curve is illustrated in Figure 4. The solid line
corresponds to the ``standard'' parameters. The dashed line
corresponds to $r_{\rm wind} = 5.0 \times 10^{18} {\rm cm}$ and the
dotted line corresponds to $r_{\rm wind} = 5.0 \times 10^{17} {\rm
cm}$. We can see that the rebrightening appears earlier when $r_{\rm
wind}$ is smaller, because $r_{\rm wind}$ determines the position of
the density jump.
\begin{figure}
\resizebox{\hsize}{!}{\includegraphics{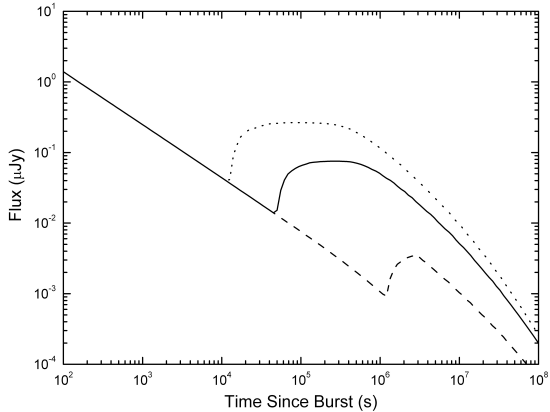}} \caption{The effect
of the parameter $r_{\rm wind}$ on the R-band light curve. The solid
line corresponds to the ``standard'' parameters. The dashed line
corresponds to $r_{\rm wind} = 5.0 \times 10^{18} {\rm cm}$ and the
dotted line corresponds to $r_{\rm wind} = 50 \times 10^{17} {\rm
cm}$.}
\end{figure}

The effect of the parameter $\alpha1$ on the R-band light curve is
shown in Figure 5. The solid line corresponds to the ``standard''
parameters. The dashed line corresponds to $\alpha1 = 1.7$ and the
dotted line corresponds to $\alpha1 = 1.3$. It is obvious that a
larger $\alpha1$ makes the flux before the rebrightening lower. This
is not difficult to understand. With the increase of $\alpha1$, the
values of $\xi_e$ and $\xi_B$ before the rebrightening become
smaller. This suppresses the radiation flux at early time.
\begin{figure}
\resizebox{\hsize}{!}{\includegraphics{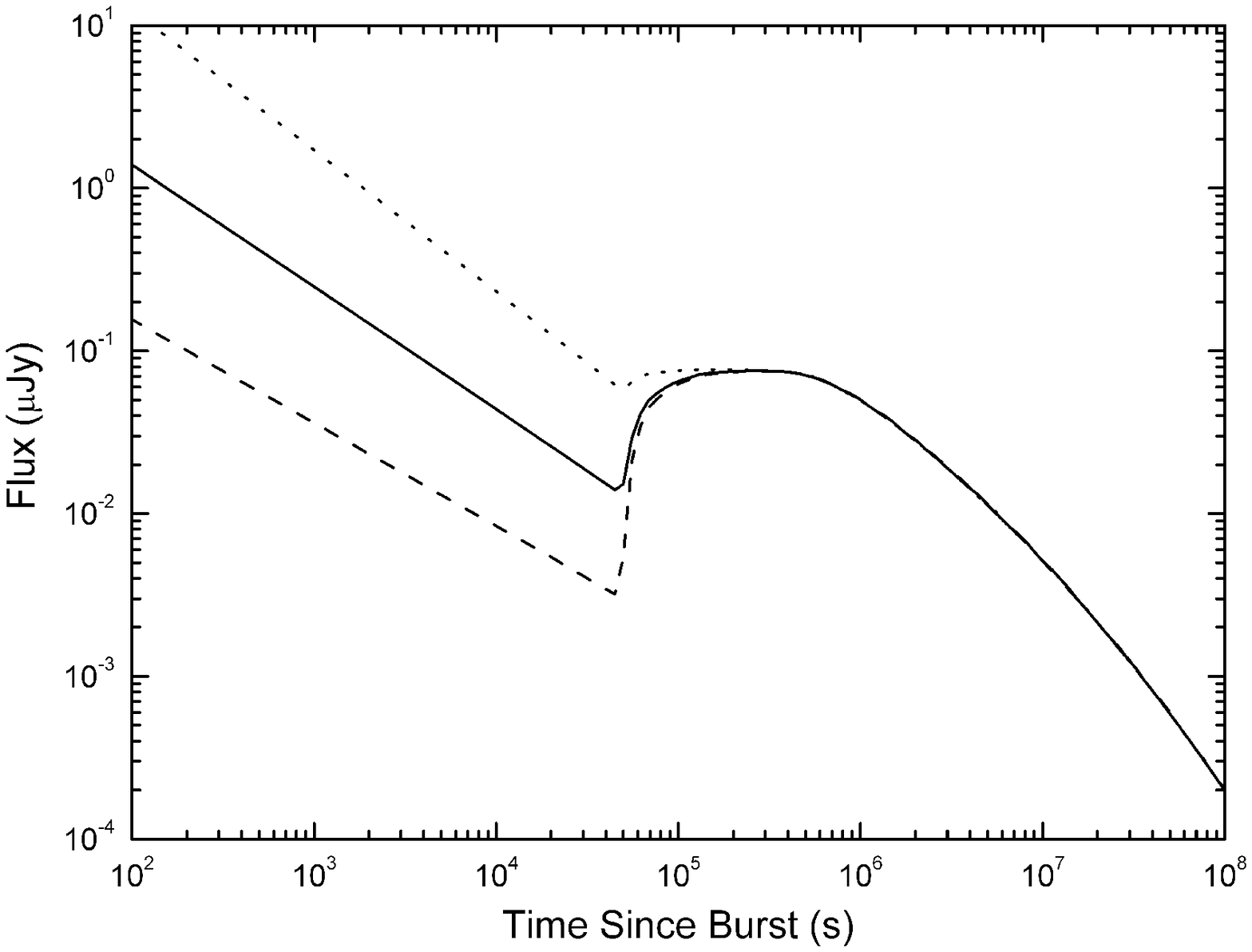}} \caption{The effect
of the parameter $\alpha1$ on the R-band light curve. The solid
line corresponds to the ``standard'' parameters. The dashed line
corresponds to $\alpha1 = 1.7$ and the dotted line corresponds to
$\alpha1 = 1.3$.}
\end{figure}

The effect of the parameter $\alpha2$ on the R-band light curve is
illustrated in Figure 6. The solid line corresponds to the
``standard'' parameters. The dashed line corresponds to $\alpha2 =
1.5$ and the dotted line corresponds to $\alpha2 = 1.1$. As
expected, the effect of $\alpha2$ is similar to that of $\alpha1$
for the same reason.
\begin{figure}
\resizebox{\hsize}{!}{\includegraphics{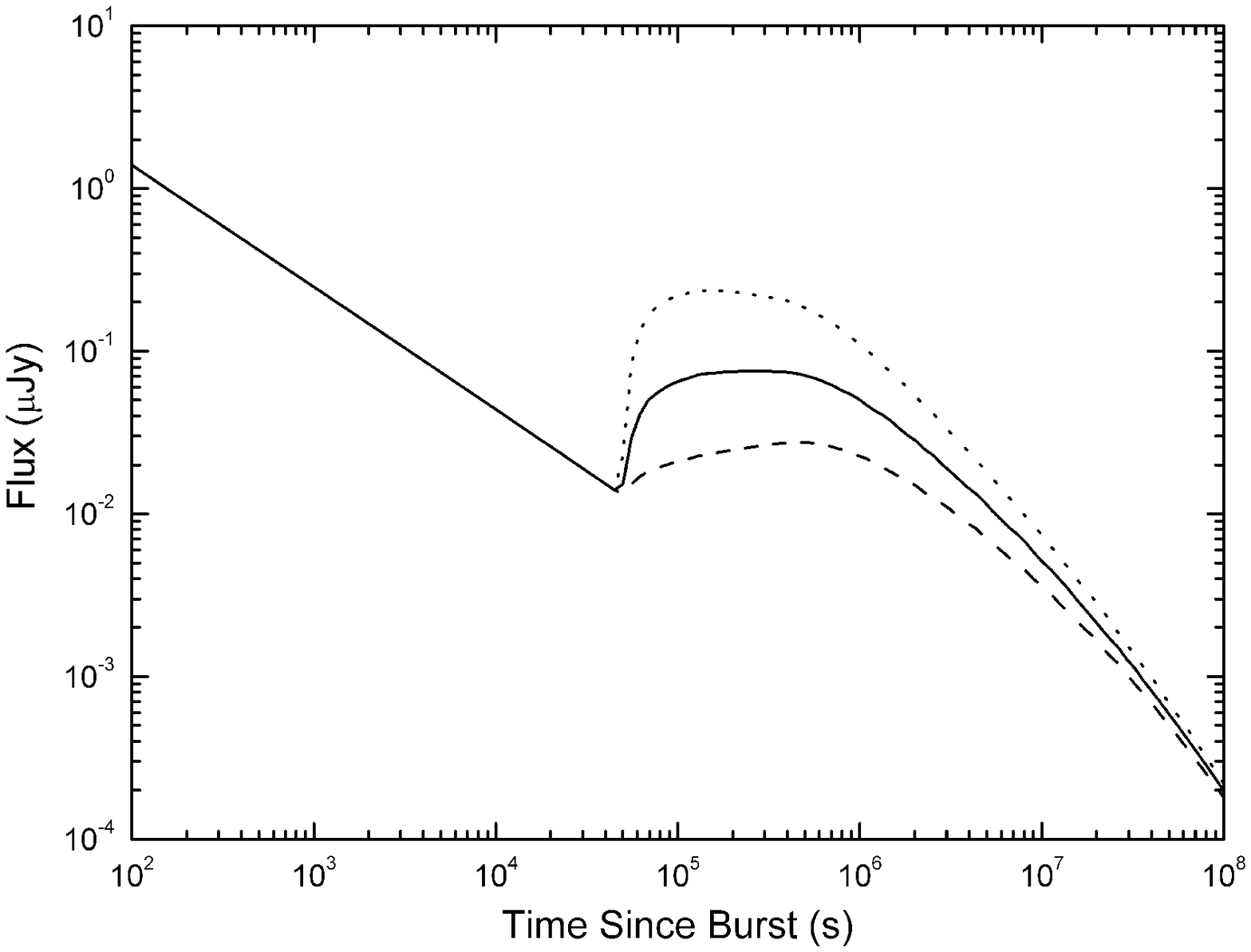}} \caption{The effect
of the parameter $\alpha2$ on the R-band light curve. The solid
line corresponds to the ``standard'' parameters. The dashed line
corresponds to $\alpha2 = 1.5.$ and the dotted line corresponds to
$\alpha2 = 1.1$.}
\end{figure}

\section{Comparison With Observations}

In this section we select three GRBs with obvious rebrightenings in
both optical and X-ray bands, i.e. GRB 060206, GRB 070311 and GRB
071010A, and reproduce their R-band and X-ray afterglow light curves
numerically, using the model described in \S 2.

\subsection{GRB 060206}

GRB 060206 was detected at 04:46:53 UT on Feb. 6 by the Burst Alert
Telescope (BAT) onboard the \emph{Swift} satellite (Morris et al.
2006). The temporal profile of the BAT light curve shows a single
peak, with a total fluence of $8.4 \pm 0.4 \times 10^{-7}$ erg ${\rm
cm}^{-2}$ in the 15-350 keV band (Morris et al. 2006; Palmer et al.
2006). The total duration of the burst is $T_{90} = 7 \pm 2 {\rm
s}$, suggesting that it is a long-duration GRB (Palmer et al. 2006).

The UVOT telescope onboard the \emph{Swift} satellite first observed
the optical afterglow of GRB 060206 and found it located at the
position of $\alpha = 13^h31^m43^s.416$, $\delta =
+35^{\circ}03'03''.6$ (J2000.0; Boyd et al. 2006), corresponding to
the Galactic extinction of $A_R = 0.033$ mag (Schlegel, Finkbeiner
\& Davis 1998). After that, a number of telescopes made detailed
follow up observations, such as the 2-m robotic Liverpool Telescope
(Monfardini et al. 2006), the Rapid Telescopes for Optical Response
(RAPTOR; Wo\'zniak et al. 2006) and the MDM telescope (Stanek et al.
2007). From the calibrated ALFOSC spectrum of the afterglow, Fynbo
et al. (2006) determined the redshift of GRB 060206 as $z = 4.045$.
X-ray afterglow was also detected by XRT on \emph{Swift} at 58 s
after the trigger (Morris et al. 2006).

There are significant rebrightenings in both the optical and X-ray
afterglows of GRB 060206 at about 3000 s after the burst. Liu et al.
(2008) suggested that the rebrightening comes from a jet-like ejecta
with a larger viewing angle, which is produced by the central
engine.

Here we try to use our model described in \S 2 to reproduce the
rebrightenings in both R-band and X-ray light curves of GRB 060206.
Our best-fit physical parameters are presented in Table 1, and our
modeling curves are illustrated in Figure 7. The observed R-band
data are taken from Monfardini et al. (2006), Wo\'zniak et al. 2006
and Stanek et al. (2007), and the X-ray data are taken from the
\emph{Swift} XRT light curve repository\footnote{$\rm
http://www.swift.ac.uk/xrt_-curves/$} (Evans et al. 2007). We can
see in Figure 7 that our model can generally reproduce the light
curves well. Note that at the late stage, the observed X-ray light
curve is obviously too flat as compared with our theoretical result.
This may be due to the contamination from a nearby source, as
already mentioned by Stanek et al. (2007).

\begin{figure}
\resizebox{\hsize}{!}{\includegraphics{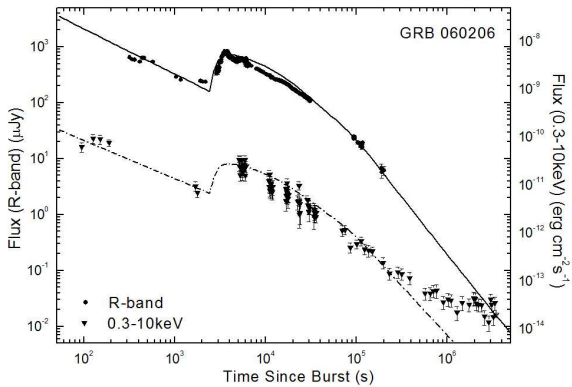}} \caption{Our best
fit to the R-band and X-ray afterglow light curves of GRB 060206.
The lines are our theoretical curves and the fitting parameter
values are given in Table 1. The R-band observed data are taken from
Monfardini et al. (2006), (Wo\'zniak et al. 2006) and Stanek et al.
(2007), and the X-ray data are taken from the \emph{Swift} XRT light
curve repository (Evans et al. 2007). All optical data points have
been corrected for the Galactic extinction (Schlegel, Finkbeiner \&
Davis 1998).}
\end{figure}

\subsection{GRB 070311}

GRB 070311 occurred at 01:52:50 UT on 2007 March 11 and was detected
by the INTEGRAL/IBAS in IBIS/ISGRI data. The duration of the
$\gamma$-ray prompt emission is about 50 s in the 20-200 keV band,
with the peak flux of 0.9 photons ${\rm cm}^{-2}$ ${\rm s}^{-1}$ (1
s integration time) and the total fluence of $2 \times 10^{-6}$ erg
${\rm cm}^{-2}$ (Mereghetti et al. 2007).

The afterglow of GRB 070311 was first discovered by the Rapid Eye
Mount (REM) telescope at 55 s after the burst. A bright fading
source was found at $\alpha = 05^h50^m08^s.21$, $\delta =
+03^{\circ}22'30''.3$ (J2000.0; Covino et al. 2007). This location
has a low Galactic latitude ($l = 202^{\circ}.766, b =
-11^{\circ}.998$ in Galactic coordinates), so the Galactic
extinction is as much as $A_R = 2.038$ mag (Schlegel, Finkbeiner \&
Davis 1998). X-ray afterglow was also found by the observations of
XRT on \emph{Swift} at about 7000 s after the trigger (Guidorzi et
al., 2007b).

The optical afterglow of GRB 070311 has complex structures. Besides
a normal power law decay, there are two fast rise exponential decay
(FRED) shaped pulses peaking around 80 and 140 s after the peak of
the GRB and possibly accompanied by the tail of prompt $\gamma$-ray
emission (Guidorzi et al. 2007a). Another structure, which is the
most attractive feature, is a significant late rebrightening between
$3 \times 10^4$ s and $2 \times 10^5$ s after the trigger in both
optical and X-ray bands (Guidorzi et al. 2007a). Guidorzi et al.
(2007a) have used a power law function plus a FRED shaped pulse to
fit the afterglows in both R and X-ray bands. More physically, they
suggested that the late afterglow rebrightening of GRB 070311 can
come from a refreshed shock. Additionally, they argued that the
density jump in the surrounding medium can be the origin of the
rebrightening too, but less appealing.

Here we use our model described in \S 2 to reproduce the late
rebrightenings in both R-band and X-ray light curves of GRB 070311.
We take the observed R-band data from Guidorzi et al. (2007a; for
REM data) and GCN circulars (Cenko 2007; Dai et al. 2007; Garnavich
et al. 2007; Halpern \& Armstrong 2007a,b,c,d; Jel\'inek, Kub\'anek
\& Prouza 2007; Kann 2007; Wren et al. 2007), and the X-ray data
from the \emph{Swift} XRT light curve repository (Evans et al.
2007). Because the redshift of GRB 070311 is still unknown, we
simply assume a redshift of $z = 1.0$ in our calculation. Our
best-fit physical parameters are presented in Table 1, and our
modeling curves are illustrated in Figure 8. We see that the
interpretation of our model to this event is also generally
acceptable.

\begin{figure}
\resizebox{\hsize}{!}{\includegraphics{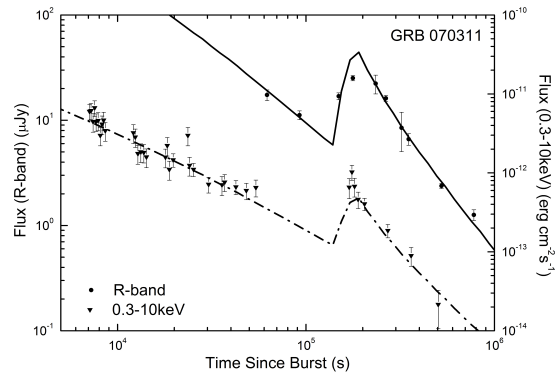}} \caption{Our best
fit to the late rebrightenings in the R-band and X-ray light curves
of GRB 070311. The lines are our theoretical curves and the fitting
parameter values are given in Table 1. The R-band observed data are
taken from Guidorzi et al. (2007a; for REM data) and GCN circulars
(Cenko 2007; Dai et al. 2007; Garnavich et al. 2007; Halpern \&
Armstrong 2007a,b,c,d; Jel\'inek, Kub\'anek \& Prouza 2007; Kann
2007; Wren et al. 2007), and the X-ray data are taken from the
\emph{Swift} XRT light curve repository (Evans et al. 2007). All
optical data points have been corrected for the Galactic extinction
(Schlegel, Finkbeiner \& Davis 1998).}
\end{figure}

\subsection{GRB 071010A}

GRB 071010A triggered the \emph{Swift} satellite at 03:41:12 UT on
2007 Oct. 10 (Moretti et al. 2007). It is a traditional long GRB
with the burst duration of $T_{90} = 6 \pm 1 {\rm s}$. The total
fluence of the burst is about $2.0 \times 10^{-7}$ erg $ {\rm
cm^{-2}}$ (Krimm et al. 2007).

TAROT telescope identified the optical afterglow of GRB 071010A
firstly, only 124 s after the trigger (Klotz, Boer \& Atteia, 2007).
Bloom et al. (2007) further found that the position of the burst was
$\alpha = 19^h12^m14^s.624$, $\delta = -32^{\circ}24'07''.16$
(J2000.0), corresponding to the Galactic extinction of $A_R = 0.263$
mag (Schlegel, Finkbeiner \& Davis 1998). Spectral observations were
done subsequently by Keck and the analysis of the Mg II and Fe II
lines gave the redshift of the burst as $z = 0.98$ (Prochaska et al.
2007). X-ray data were also acquired by XRT observations on
\emph{Swift} (Guidorzi et al. 2007b), but no radio afterglow was
detected at 8.46 GHz until almost 2 days after the burst (Chandra \&
Frail 2007).

The most interesting feature of the afterglows of GRB 071010A is the
appearance of a sharp rebrightening at about 0.6 days after the GRB,
clearly seen in both optical and X-ray bands. Covino et al. (2008)
have used a Beuermann et al. (1999) function, and a simple step
function for the rebrightening, to fit the afterglows. They
suggested that the rebrightening was due to an injection of abundant
energy, which was comparable to the initial energy in the fireball.
The steepening after the rebrightening was interpreted as a jet
break at around 1 day after the burst.

We try to use our model introduced in \S 2 to reproduce the R-band
and X-ray light curves of GRB 071010A. In our modeling, the observed
R-band data are taken from Covino et al. (2008), and the X-ray data
are taken from the \emph{Swift} XRT light curve repository (Evans et
al. 2007). Our best-fitting physical parameters are presented in
Table 1, and our modeling curves are illustrated in Figure 9. It is
clear that our model can reproduce the rebrightening features of GRB
071010A very well.

\begin{figure}
\resizebox{\hsize}{!}{\includegraphics{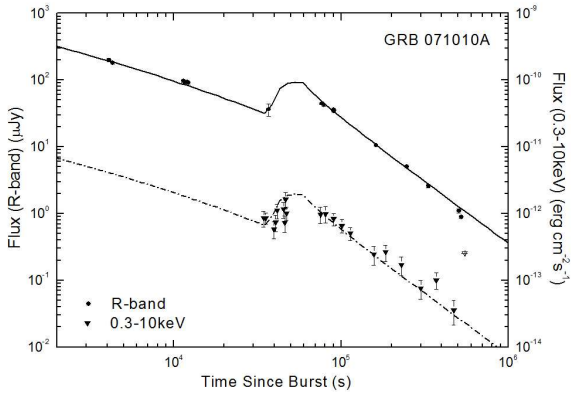}} \caption{Our best
fit to the R-band and X-ray afterglow light curves of GRB 071010A.
The lines are our theoretical curves and the fitting parameter
values are given in Table 1. The R-band observed data are taken from
Covino et al. (2008), and the X-ray data are taken from the
\emph{Swift} XRT light curve repository (Evans et al. 2007). All
data points have been corrected for the Galactic extinction
(Schlegel, Finkbeiner \& Davis 1998).}
\end{figure}

\section{Discussion and Conclusion}

In the standard fireball model, it is assumed that the circum-burst
medium density is constant or following a single $\rho
\propto r^{-2}$ law. The microphysics shock parameters are also
usually assumed invariable during the evolution of the fireball.
These models can basically explaining many pre-$Swift$
GRB afterglows, whose spectrum and
light curves can be approximated as broken power-law functions
(Panaitescu \& Kumar 2001a, 2001b; Yost et al. 2003). The launch of
\emph{Swift} satellite (Gehrels 2004) makes it possible to
observe early afterglows of many GRBs in the first few hours after the
trigger. Many remarkable and unexpected features are found,
such as marked rebrightenings and the flattish decay phase in early GRB
afterglows.

In fact, the standard fireball model is obviously too simplified. If
long GRBs indeed originate from the death of massive stars, the
environment of GRBs should have complex structures rather than have
a constant or single power law profile. Some analytical (Castor et
al. 1975; Weaver 1977; Ramirez-Riuz et al. 2005; Pe'er and Wijers
2006) and numerical (Ramirez-Riuz et al. 2001) studies suggest that
the circum-burst density profile should be a wind bubble, associated
with a few density jumps. On the other hand, the microphysics
parameters, such as $\xi_{\rm e}$ and $\xi_{\rm B}$, may vary during
the evolution of the fireball (Rossi \& Rees 2003; Ioka et al. 2006;
Fan \& Piran 2006; Panaitescu et al. 2006; Granot, K\"{o}nigl \&
Piran 2006). Actually, the fast decrease of the cooling frequency
$\nu_c$ in some GRBs suggests that $\xi_{\rm B}$ may be evolving
(Panaitescu et al. 2006). Some previous studies also suggest that
$\xi_{\rm e}$ and $\xi_{\rm B}$ may be different for the forward
shock and the reverse shock (Fan et al. 2002; Wei, Yan \& Fan 2006),
and may also be different for Region (1) and Region (2) (Gebdre et
al. 2007; Kamble, Resmi \& Misra 2007). So, $\xi_{\rm e}$ and
$\xi_{\rm B}$ may depend on the strength of the shock and the
environment.

In our study, we combine the wind bubble environments and the change
of microphysics shock parameters together. Comparing with the
standard fireball model, our model has three more parameters (i.e.
$r_{\rm wind}$, $\alpha1$ and $\alpha2$).  We find that this model
can produce the observed rebrightenings and flattish decay in GRB
afterglows successfully. We have shown that the observed
rebrightenings in both the optical and X-ray afterglow light curves
of GRB 060206, GRB 070311 and GRB 071010A can be well explained by
our model.

We have also investigated the effects of various parameters on the
light curves numerically. We can imagine that if the
stellar wind produced by the progenitor is very strong, or the
launching speed of the stellar wind $v_{\rm w}$ is small, or
the value of $r_{\rm wind}$ is large enough, there will be no
rebrightening within the usual observation time.
So, basically the wind bubble environment can also give birth to
a steadily decaying afterglow that shows no rebrightening.

In our work, we neglect the effect of the reverse shock.
According to the study by Dai \& Lu (2002),
when an ultrarelativistic blast wave interacts with a density jump
medium, the resulting reverse shock is relativistic only if the
amplitude of the density jump is much larger than 21. In our model, the
amplitude of the density jump is only 4, so the corresponding reverse
shock is Newtonian. The emission from the Newtonian reverse shock is
very weak, and can be omitted.
Moreover, although we use an abrupt density jump, the actual increase
of density may be gradual. In this situation, the emission
from the reverse shock will even be much weaker.

Currently, a complete understanding of the microphysical processes
in the relativistic shocks is still lacking. Using the derived
microphysical parameters from GRB modeling, people may be able to
get some constraints on the shock physics. The derived parameter
values of $\dot M$ and $n_{\rm ISM}$ are also very important. They
are closely related to the evolution and the environment of the
massive star. They may give some hints on the characters of the
progenitor, such as the initial main-sequence mass and the
metallicity. So, from the modeling parameters, we can also obtain
some useful information about GRB origin.

\section*{Acknowledgments}

We thank the anonymous referee for stimulating suggestions that lead
to an overall improvement of this study. We also would like to thank
Kinwah Wu, Z. G. Dai and X. Y. Wang for helpful suggestions and
discussion. This research was supported by a 2009 GRF grant of Hong
Kong Government (grants 701109), the National Natural Science
Foundation of China (grants 10625313), and the National Basic
Research Program of China (973 Program 2009CB824800).

\label{lastpage}

\end{document}